\documentclass{article}
\usepackage{spconf,amsmath,graphicx}
\usepackage{arydshln}
\usepackage{xcolor}
\usepackage{slashbox}
\usepackage{multirow}
\usepackage{amsfonts}
\usepackage{citesort}

\title{A COMPARATIVE STUDY OF MODULAR AND JOINT APPROACHES FOR SPEAKER-ATTRIBUTED ASR ON MONAURAL LONG-FORM AUDIO}
\name{\begin{tabular}{c}Naoyuki Kanda, Xiong Xiao, Jian Wu, Tianyan Zhou, Yashesh Gaur, Xiaofei Wang, Zhong Meng, \\Zhuo Chen, Takuya Yoshioka 
\end{tabular}}
\address{Microsoft Corp., USA}

\begin{document}
\ninept
\maketitle
\begin{abstract}
Speaker-attributed automatic speech recognition (SA-ASR) is a task to recognize ``who spoke what'' from multi-talker recordings. An SA-ASR system usually consists of multiple modules such as speech separation, speaker diarization and ASR. On the other hand, considering the joint optimization, an end-to-end (E2E) SA-ASR model has recently been proposed with promising results on simulation data. In this paper, we present our recent study on the comparison of such modular and joint approaches towards SA-ASR on real monaural recordings. We develop state-of-the-art SA-ASR systems for both modular and joint approaches by leveraging large-scale training data, including 75 thousand hours of ASR training data and the VoxCeleb corpus for speaker representation learning. We also propose a new pipeline that performs the E2E SA-ASR model after speaker clustering. Our evaluation on the AMI meeting corpus reveals that after fine-tuning with a small real data, the joint system performs 8.9--29.9\% better in accuracy compared to the best modular system while the modular system performs better before such fine-tuning. We also conduct various error analyses to show the remaining issues for the monaural SA-ASR.
\end{abstract}
\section{Introduction}
\label{sec:intro}

Speaker-attributed automatic speech recognition (SA-ASR)
 is a task to recognize 
``who spoke what''
from multi-talker recordings.
It has been long studied toward meeting and conversation analysis 
from the research project in 2000s \cite{janin2003icsi,carletta2005ami,fiscus2007rich} 
to the recent international competition
such as CHiME-5/6 Challenges \cite{barker2018fifth,watanabe2020chime}.
An SA-ASR system usually consists of
multiple modules such as speech separation to handle overlapping speech, 
 speaker diarization to count and assign the speaker information,
and ASR to transcribe
the contents.
There have been a lot of studies on microphone array recordings
 to improve speech separation \cite{anguera2007acoustic,yoshioka2019advances,kanda2019guided}, 
 speaker diarization \cite{anguera2007acoustic,medennikov2020target}
 and ASR systems \cite{xiao2016deep,kanda2019acoustic} 
 by using spatial information.
On the other hand, SA-ASR based on 
 a single microphone is still highly challenging,
 and only limited amount of studies have been conducted
 for a fully automatic SA-ASR system 
 on the monaural long-form audio \cite{mao2020speech,raj2020integration,kanda2020minimum}.

One potential problem that limits the accuracy of an SA-ASR system
is its difficulty to optimize multiple modules in the SA-ASR system.
For example,
speech separation models are usually trained based on the signal-level criterion,
but it is not necessarily optimal for ASR or speaker diarization. 
To mitigate such sub-optimality,
there has been a series of studies for a joint model that
combines multiple modules such as 
joint speech separation and ASR \cite{yu2017recognizing,seki2018purely,chang2019end,chang2019mimo,kanda2019acoustic,kanda2019auxiliary},
joint speaker identification/diarization and speech separation \cite{wang2019speech,von2019all,kinoshita2020tackling},
% joint speaker diarization and speech separation \cite{}
 or joint speech recognition and speaker diarization \cite{el2019joint,kanda2019simultaneous,mao2020speech}.
Recently, 
an end-to-end (E2E) SA-ASR model that jointly performs speaker counting,
multi-talker speech recognition, and speaker identification
was proposed with a promising result for simulation data \cite{kanda2020joint}.
For LibriCSS data \cite{chen2020continuous}, 
which is a playback of LibriSpeech corpus \cite{panayotov2015librispeech}
 in a real meeting room, the monaural E2E SA-ASR model achieved even better 
 accuracy than the modular system with multi-channel separation.

While promising results were shown for such joint systems, 
most of the previous studies 
were limited to 
either simulated data \cite{yu2017recognizing,chang2019end,zhang2020improving,chang2020end,tripathi2020end,kanda2020sot,kanda2020joint,kanda2020investigation,lu2021streaming,sklyar2021streaming,chang2021hypothesis} or small-scale real data 
\cite{ko2017study,ganapathy20183,peddinti2017low,kanda2019acoustic}.
It is because of the scarcity of training data for real meeting recordings,
which takes a lot of time to precisely transcribe.
Recently, it is shown that 
a simulation-based large-scale pre-training of multi-talker ASR (i.e. joint model of speech separation
and ASR) followed by fine-tuning with a small real data is very effective for
real meeting transcription \cite{kanda2021large}.
However, it is still unclear how we can leverage such a pre-training scheme
for the E2E SA-ASR model that includes speaker identification.
For example, 
speaker information is often missing in ASR training data
while 
the reference transcription is often missing in training data for speaker representation learning.
%ASR training data is often missing speaker information
%a training data for speaker representation learning 
%is often missing the reference transcription.
Therefore,
even a simple simulation of 
multi-talker audio with reference transcription and speaker information is not easy
although it is required to pre-train the E2E SA-ASR model.

With these background,
in this paper,
we present our recent study on the comparison of 
modular and joint approaches for the SA-ASR system with
real meeting recordings
from a single microphone.
To better handle complicated acoustic and linguistic characteristics
of real data,
we fully leverage large-scale training data 
that include the 75 thousand (K) hours of our internal ASR training data
and the VoxCeleb corpus \cite{nagrani2017voxceleb,chung2018voxceleb2} for speaker representation learning.
For the pre-training of the E2E SA-ASR model,
we develop a new pre-training scheme that partly uses pseudo labels of
training data.
We also propose a new pipeline that performs the E2E SA-ASR model after conventional clustering-based
speaker diarization.
We show various comparison and error analyses based on the AMI meeting corpus \cite{carletta2005ami}, 
which shows the benefit of the joint approach as well as the remaining issues for the monaural SA-ASR.

\section{Evaluation Data and Metric}
\label{sec:data}

In this study, we use the AMI meeting corpus \cite{carletta2005ami} 
for the evaluation of various SA-ASR pipelines.
The corpus comprises approximately 100 hours of meeting recordings.
Most meetings consist of four participants while some recordings contain
speech from three or five participants.
The audio
was recorded by an 8-ch microphone array, which is often called 
multiple distant microphone ({\bf MDM}).
The first channel of the MDM audio is used for monaural evaluation,
referred to as a single distant microphone ({\bf SDM}) setting.
The AMI corpus
also contains the recordings of the same meetings from independent headset microphones ({\bf IHM})
worn by each participant.

The primary focus of this paper is the evaluation on SDM recordings.
Meanwhile, we 
also evaluated the audio that is generated by simply mixing IHM recordings of all participants,
which is referred to as {\bf IHM-MIX} audio.
 We used scripts in Kaldi toolkit \cite{povey2011kaldi} to
partition
the recordings into training, development and evaluation sets.
The total duration for the three sets is
  80.2 hours, 9.7 hours, and 9.1 hours, respectively.
We tuned the system parameters based on the development set of the SDM recordings.

In the evaluation, we used
 the concatenated minimum-permutation
word error rate (cpWER) \cite{watanabe2020chime} as a primary metric.
The cpWER is computed as follows:
(i) concatenate all reference transcriptions in chronological order for each speaker; 
(ii) concatenate all hypothesis transcriptions in  chronological order for each detected speaker; 
(iii) compute the WER between the concatenated references and concatenated hypotheses for all possible
speaker permutations, and pick the lowest WER among them.
The cpWER is affected by both
the speech recognition and speaker diarization results.
Besides the cpWER, we used
speaker counting error (SCE), defined as
the per-meeting average of absolute difference between the estimated number of speakers
and the actual number of speakers, and 
word error rate (WER)
%, and 
%diarization error rate (DER) with 0.25 sec of error tolerance and with overlapping segment,
on demand in the error analysis.

\section{E2E SA-ASR: Review}
\label{sec:e2e_sa_asr}

\begin{figure}[t]
  \centering
  \includegraphics[width=\linewidth]{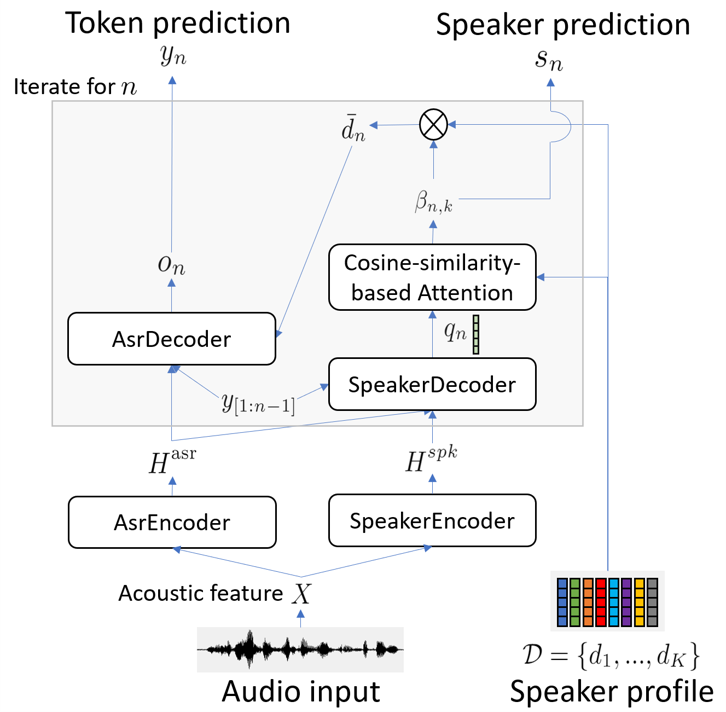}
%  \vspace{-7mm}
  \caption{E2E SA-ASR model.}
  \label{fig:asr_models}
%  \vspace{-5mm}
\end{figure}

This section gives an overview of the E2E SA-ASR model \cite{kanda2020joint} which is the 
basis of our joint modeling pipeline.
It 
 jointly performs speaker counting, multi-talker speech recognition,
and speaker identification from potentially overlapping speech.
The overview of the E2E SA-ASR is shown in Fig. \ref{fig:asr_models}.

\subsection{Problem statement}
Suppose 
we observe a sequence of acoustic feature $X\in\mathbb{R}^{f^a\times l^a}$ extracted from audio input that possibly includes overlapping speech, 
where $f^a$ and $l^a$ are the feature dimension and the length of the feature sequence,
respectively.
Also suppose 
that we have a set of 
 speaker profiles $D=\{d_k \in \mathbb{R}^{f^d}|k=1,...,K\}$,
where $K$ is the total number of profiles, 
 $d_k$ is the speaker embedding 
 (e.g., d-vector \cite{variani2014deep})
of the $k$-th speaker,
and $f^d$ is the dimension of the speaker embedding.
We assume $D$ includes profiles of the speakers in the observed audio.
$K$ can be greater than the actual number of speakers in the recording.

Given
 $X$ and $D$,
the E2E SA-ASR model %\cite{kanda2020joint}
   estimates a
  multi-speaker transcription $Y=(y_n\in \{1,...,|\mathcal{V}|\}|n=1,...,N)$
accompanied by
the speaker identity of each token $S=(s_n\in \{1,...,K\}|n=1,...,N)$. 
Here,
$|\mathcal{V}|$  is the size of the vocabulary $\mathcal{V}$, 
and  $y_n$ and $s_n$ are the word index and speaker index for the $n$-th token,
 respectively.
  Following the serialized output training (SOT) framework \cite{kanda2020sot}, we represent 
a multi-speaker transcription 
as a single sequence $Y$ by joining word sequences of multiple speakers with a special ``speaker change'' symbol $\langle sc\rangle$.
 For example, the reference token sequence to $Y$ for the three-speaker case is given as
$R=\{r^1_{1},..,r^1_{N^1}, \langle sc\rangle, r^2_{1},..,r^2_{N^2}, \langle sc\rangle, r^3_{1},..,r^3_{N^3}, \langle eos\rangle\}$, 
where $r^j_i$ represents the $i$-th token of the $j$-th speaker.
A special symbol $\langle eos\rangle$ is inserted at the end of all transcriptions.
Note that this representation of multi-speaker transcription
can be used even for overlapping speech.
The order of speakers in the reference transcription is determined by
the start time of the utterance of each speaker, a.k.a. first-in, first-out SOT \cite{kanda2020sot}.

\begin{figure*}[t]
  \centering
  \includegraphics[width=\linewidth]{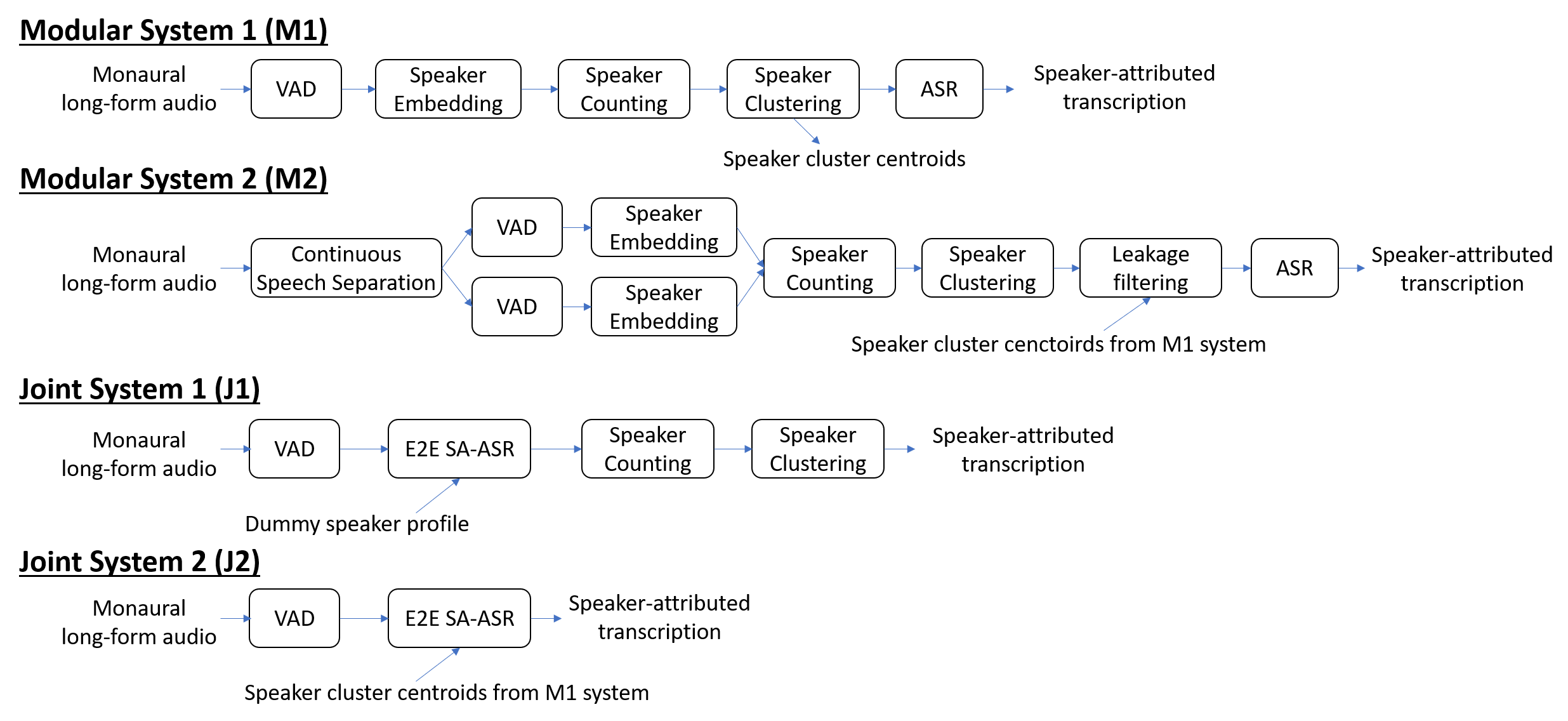}
  \vspace{-7mm}
  \caption{Modular and joint systems for SA-ASR on monaural long-form audio.}
  \label{fig:systems}
%  \vspace{-5mm}
\end{figure*}

\subsection{Model architecture}

The E2E SA-ASR model consists of 
the {\it ASR block} and the {\it speaker block},
which jointly perform
ASR and speaker identification.
The ASR block follows an attention-based encoder-decoder (AED) design 
and is 
represented as, 
 \begin{align}
 H^{\rm asr} &={\rm AsrEncoder}(X),  \label{eq:enc}  \\
 o_n &= {\rm AsrDecoder}(y_{[1:n-1]}, H^{\rm asr}, \bar{d}_n).  \label{eq:asrout}
 \end{align}
%Given the acoustic feature $X$, 
An AsrEncoder module
first converts the acoustic feature $X$ 
into a sequence of hidden embeddings $H^{\rm asr} \in \mathbb{R}^{f^h\times l^h}$ for ASR (Eq. \eqref{eq:enc}),
where $f^h$ and $l^h$ are the embedding dimension and 
the length of the embedding sequence, respectively.
An AsrDecoder module then iteratively estimates $Y$ for $n=1,...,N$.
At each decoder step $n$, 
the AsrDecoder % module 
calculates
the output distribution $o_n \in \mathbb{R}^{|\mathcal{V}|}$ 
 given
previous token estimates $y_{[1:n-1]}$,
$H^{\rm asr}$,
and 
the weighted speaker profile $\bar{d}_n$ (Eq. \eqref{eq:asrout}).
Here, $\bar{d}_n$ is 
the weighted average of the speaker profiles $D$
calculated in the speaker block, 
which will be
explained later.
The posterior probability
of token $i$ (i.e. the $i$-th token in $\mathcal{V}$) 
at the $n$-th decoder step 
is represented as
 \begin{align}
Pr(y_n=i|y_{[1:n-1]},s_{[1:n]},X,D) = o_{n,i}, \label{eq:tokenprob}
\end{align}
where $o_{n,i}$ represents
the $i$-th element of $o_n$.

On the other hand, the speaker block is also designed based on the AED architecture as % follows.
\begin{align}
 H^{spk} &= {\rm SpeakerEncoder}(X),  \label{eq:spkenc} \\
 q_n &= {\rm SpeakerDecoder}(y_{[1:n-1]},H^{\rm spk},H^{\rm asr}), \label{eq:spkquery} \\
\beta_{n,k}&= \frac{\exp(\cos(q_n,d_k))}{\sum_j^K \exp(\cos(q_n,d_j))}, \label{eq:invatt} \\
\bar{d}_n&=\sum_{k=1}^{K}\beta_{n,k}d_k. \label{eq:weighted_prof}
\end{align}
A SpeakerEncoder module first converts $X$ into 
a sequence of speaker embeddings $H^{\rm spk}\in \mathbb{R}^{f^h \times l^h}$
that represents the speaker characteristic of
 $X$ (Eq. \eqref{eq:spkenc}).
A SpeakerDecoder module then 
iteratively estimates $S$ for $n=1,...,N$.
At every decoder step $n$,
the SpeakerDecoder
calculates
a speaker query $q_n \in \mathbb{R}^{f^d}$ given $y_{[1:n-1]}$,
$H^{\rm spk}$ and $H^{\rm asr}$ (Eq. \eqref{eq:spkquery}).
Next,
a cosine similarity-based attention weight
 $\beta_{n,k}\in \mathbb{R}$
is calculated 
for all 
 profiles $d_k$ in $D$
given the speaker query $q_n$ (Eq. \eqref{eq:invatt}).
A posterior probability of person $k$ speaking the $n$-th token given all the previous estimation and observation
%as well as the inputs $X$ and $D$ 
is represented by $\beta_{n,k}$ as
\begin{align}
Pr(s_n=k|y_{[1:n-1]},s_{[1:n-1]},X,D)=\beta_{n,k}. \label{eq:spk-prob}
\end{align}
Finally,
a weighted average of the speaker profiles is calculated as $\bar{d}_n\in \mathbb{R}^{f^d}$ (Eq. \eqref{eq:weighted_prof}) to be fed into the ASR block 
 (Eq. \eqref{eq:asrout}).

The joint posterior probability of token $Y$ and speaker $S$ given $X$ and $D$
can be represented based on Eqs. \eqref{eq:tokenprob} and \eqref{eq:spk-prob} as follows. 
\begin{align}
Pr(Y,S|X,D) =&\prod_{n=1}^{N}\{Pr(y_{n}|y_{[1:n-1]}, s_{[1:n]}, X, D) \nonumber \\ 
&\;\;\times Pr(s_{n}|y_{[1:n-1]}, s_{[1:n-1]}, X, D) \}. \label{eq:samll-2}
\end{align}
The model parameters of the E2E SA-ASR are optimized by maximizing $Pr(Y,S|X,D)$ over
training data.

In this paper, we 
use Transformer-based network architecture 
for the AsrEncoder, AsrDecoder and SpeakerDecoder modules,
whose details are described in \cite{kanda2021end}.
The network architecture of the SpeakerEncoder module 
is based on Res2Net \cite{gao2019res2net}
and designed to be the same with that of the speaker profile extractor 
except the final average-pooling layer.

\section{Modular and joint systems for speaker-attributed ASR}
This section explains four SA-ASR systems that we investigate in our evaluation.
The overview of each SA-ASR system is shown in Fig. \ref{fig:systems}.
Note that the system named ``Joint System 2 (J2)'' is a new pipeline that we propose in this paper 
while other three systems are known pipelines for SA-ASR that has been investigated in prior works (such as \cite{raj2020integration,kanda2020investigation}).

\subsection{Modular systems}
\subsubsection{Modular system 1 (M1)}
The modular system 1 (M1) is a system that combines conventional clustering-based speaker diarization
with a single-talker ASR.
The M1 system can not handle overlapping speech appropriately
due to the limitation of the conventional clustering-based speaker diarization \cite{park2021review}.

In the M1 system, monaural long-form audio is first divided into multiple short segments based on 
the voice activity detector (VAD). 
Specifically, WebRTC VAD\footnote{https://github.com/wiseman/py-webrtcvad} is used
with a least aggressive setting to
keep speech region as much as possible.
Then,
128-dim d-vector is extracted for every 0.75 sec
with 1.5 sec of window.
For d-vector extraction, we use Res2Net-based network trained with VoxCeleb corpora \cite{nagrani2017voxceleb,chung2018voxceleb2},
which is the same one used in our speaker diarization system \cite{xiao2021microsoft} that won 
the VoxSRC2020 Challenge \cite{nagrani2020voxsrc}.
After extracting the d-vector, we applied normalized maximum eigengap (NME)-based speaker counting \cite{park2019auto}.
Spectral clustering is then applied based on the estimated number of speakers by NME.
Finally, a single-talker ASR system is applied based on the speaker diarization result.
When we apply the ASR, we concatenate the speech regions that have less than 1 sec of intermediate silence 
as long as the total duration becomes less than 20 sec. If the speech region is longer than 20 sec,
such speech region is simply split by 20 sec of sliding window without overlap.

\subsubsection{Modular system 2 (M2)}
The modular system 2 (M2) introduces the continuous speech separation (CSS) module \cite{yoshioka2019advances,chen2020continuous,chen2021continuous}
to handle overlapping speech.
It also introduces the leakage filtering \cite{xiao2021microsoft} to
remove residual noise that is sometimes generated by the CSS module.

In the M2 system,
the CSS is first applied to convert the possibly overlapping audio into
two streams of continuous single-talker audio.
Here, we assume that the number of simultaneously overlapping speakers is up to two.
We use the same CSS model that was used in our speaker diarization system \cite{xiao2021microsoft} 
that won the 
VoxSRC2020 Challenge.
It consists of 
18 Conformer encoder layers with 8 attention heads, 512 attention dimensions and 
1024 feed-forward network dimension,
and was trained by 1500 hours of simulated  mixed training samples.
After the CSS module, we apply the VAD and speaker embedding extraction for both
speech streams
with the same setting as in the M1 system.
The speaker embeddings from two streams are pooled,
and the speaker counting by NME and speaker clustering by the spectral clustering are 
applied.
Then, the leakage filtering procedure is applied to exclude the speaker cluster
that comprises only noise signal. 
Specifically,
for each speaker cluster estimated by the M2 system,
we calculate the cosine distance between
the centroid of the cluster with all cluster centroids that are estimated by the M1 system.
If the minimum cosine distance is larger than the pre-defined threshold of 0.05,
we exclude such a cluster by regarding it as a noise-only cluster.
Finally, we apply the single-talker ASR with the same setting as in the M1 system.

\subsubsection{Details of the single-talker ASR}
The single talker ASR was developed based on the AED.
The encoder consisted of 
2 layers of convolution layers that subsamples the time frame by a factor of 4,
followed by
18 layers of variant of Conformer \cite{gulati2020conformer}, where we
inserted a squeeze-and-excitation (SE) module \cite{hu2018squeeze} before the dropout of the convolution module, 
removed batch normalization, and 
added one more point-wise convolution after depth-wise convolution.
Each Conformer layer consisted of
two 1024-dim feed forward layers in a sandwich structure,
a multi-head attention with 8 heads,
a depthwise convolution with kernel size 3,
and an SE network with reduction factor 8
\cite{hu2018squeeze}.
Embedding dimension was set to 512.
The decoder consisted of
6 layers, each of which had the
multi-head attention with 8 heads, and 2048-dim
feed forward layer.
We used 80-dim log mel filterbank extracted every 10 msec as the input feature.
4K subwords \cite{kudo2018subword}
were used as recognition units.

The ASR system was first pre-trained based on 
64 million 
 anonymized and transcribed English utterances,
totaling 75K hours.
The data includes audio from various domains such as voice search and dictation.
We performed 425K training iterations with
32 GPUs, each of which consumed mini-batches of 24,000 frames.
We used Adam optimizer with
a linear decay learning rate schedule with a peak learning rate of 1e-3 
after 25K warm up iterations.
After the pre-training, 
we further fine-tuned the model by using the AMI-SDM training data,
in which we included 1 hour of noise-only segments from SDM recordings.
We found such inclusion of noise-only segments was effective to reduce 
insertion errors in the inference on noise-only segments due to the failure of the VAD module. 
We conducted 2,500 training iterations with 16 GPUs, each of which
consumed mini-batches of 6,000 frames. 
A linear decay learning rate schedule starting at a learning rate of 1e-4 was used.

\subsection{Joint systems}
\subsubsection{Joint system 1 (J1)}

The joint system 1 (J1) is a system that uses the E2E SA-ASR system in the pipeline.
Because the E2E SA-ASR system requires the speaker profile as an input,
we use a trick to feed ``dummy'' speaker profiles that is originally proposed in \cite{kanda2020investigation}.
Since we cannot rely on the speaker identification result on such dummy speaker profiles,
we instead apply speaker counting and clustering based on the speaker query generated by the E2E SA-ASR model.

In the J1 system,
we first apply the WebRTC VAD with the least aggressive setting
to segment the long-form audio into multiple short segments.
Here, similar to the pre-processing before ASR in the modular systems,
we concatenate the short segments that have less than 1 sec of intermediate silence 
as long as the total duration becomes less than 20 sec. 
If the length of the segment is longer than 20 sec,
such a segment is simply split by 20 sec of sliding window without overlap. 
We then apply the E2E SA-ASR on each short segment.
Here, we feed ``dummy'' speaker profiles that are extracted from 100 random utterances
from LibriSpeech corpus \cite{panayotov2015librispeech}.
As the result of the E2E SA-ASR, we obtain the multi-talker transcription.
In addition,
we store the speaker query $q_n$ at the position of 
$\langle sc\rangle$ token or
$\langle eos\rangle$ token
for each estimated utterance.
%as a representative speaker embedding for corresponding utterance.
We apply the NME-based speaker counting followed by the spectral clustering 
on the speaker query to assign the speaker for each estimated utterance.

\subsubsection{Joint system 2 (J2)}

The joint system 2 (J2) is a system that uses the speaker centroids estimated by the M1 system.
This pipeline is evaluated for the first time in this paper.

In the J2 system, VAD is first applied to segment the long-form audio
into multiple short segments with the same configuration as in the J1 system.
Then, the E2E SA-ASR is applied for each short segments
by feeding the speaker cluster centroids estimated by the M1 system as the speaker profiles.
Here, we apply speaker deduplication method \cite{kanda2021end} in the inference by the E2E SA-ASR model.
Specifically,
the speaker estimation is conducted with 
 a constraint
that the same speaker cannot be assigned for the consecutive
utterances joined by $\langle sc\rangle$ token. 
Given that constraint, a sequence of
speakers that has the highest speaker probability 
(multiplication of $\beta_{n,k}$) among all possible speaker sequences is selected.
We simply use the estimated speaker index for each estimated utterance
 as the speaker diarization result.

\setlength{\dashlinedash}{2pt}
\setlength{\dashlinegap}{2pt}
\begin{table*}[t]
  \caption{The cpWER (\%) and SCE for various modular and joint systems for AMI meeting corpus. All systems are tuned for cpWER on the AMI SDM development set.}
  \label{tab:main}
  \centering
  {
  \begin{tabular}{c|ccc|c|ccc|ccc|c|c}
  \multicolumn{6}{c}{\it (a) AMI SDM development set.} && \multicolumn{6}{c}{\it (b) AMI SDM evaluation set.} \\
   \cline{1-6} \cline{8-13}
  System   & \multicolumn{4}{c|}{cpWER (\%)}  & SCE && System   & \multicolumn{4}{c|}{cpWER (\%)}  & SCE \\
  ID       & sub & del & ins & {\bf total} &  &&ID       & sub & del & ins & {\bf total} & \\
    \cline{1-6} \cline{8-13}
  M1  & 7.8  & 18.6  & 2.2 & {\bf 28.6}  & 0.56 & & M1 & 7.9 & 20.0 & 2.4 & {\bf 30.3} & 0.69 \\
  M2  & 9.2  & 11.6  & 4.0  & {\bf 24.8}  & 0.00 & & M2 & 10.0  & 14.3  & 4.1  & {\bf 28.4} & 0.00 \\ \cline{1-6} \cline{8-13}
  J1  & 8.5  & 10.1 & 6.4 & {\bf 25.0}  & 0.39 & & J1 & 10.1 & 11.3 & 6.2 & {\bf 27.7} & 0.63  \\
  J2  & 8.5  & 9.0  & 5.0 & {\bf 22.6}  & 0.56  & & J2 & 10.5 & 9.9 & 4.4 & {\bf 24.9} & 0.69\\ \cline{1-6} \cline{8-13}
  \multicolumn{13}{c}{}\\
      \multicolumn{6}{c}{\it (c) AMI IHM-MIX development set.} & $\;\;\;\;\;\;\;\;\;\;\;$ & \multicolumn{6}{c}{\it (d) AMI IHM-MIX evaluation set.} \\
   \cline{1-6} \cline{8-13}
  System   & \multicolumn{4}{c|}{cpWER (\%)}  & SCE & & System   & \multicolumn{4}{c|}{cpWER (\%)}  & SCE \\
  ID       & sub & del & ins & {\bf total} &    & & ID       & sub & del & ins & {\bf total} &  \\
   \cline{1-6} \cline{8-13}
  M1  & 6.0  & 14.7  & 2.4 & {\bf 23.1} &0.89  && M1 & 5.7  & 15.3 & 2.3  & {\bf 23.4}  & 0.94  \\
  M2  & 6.3  & 13.0 & 3.3  & {\bf 22.6} & 0.44  && M2 & 6.4  & 14.2  & 2.8  & {\bf 23.4}  & 0.44  \\ \cline{1-6} \cline{8-13}
  J1  &  5.9 & 7.0  & 7.9 & {\bf 20.8} & 0.67  && J1 & 6.6  & 6.0  & 5.5   & {\bf 18.0}  & 0.38  \\
  J2  & 6.1  & 4.6  & 5.2 & {\bf 15.9} & 0.89 && J2 & 6.6  & 5.3  & 4.5  & {\bf 16.4}  & 0.94 \\ \cline{1-6} \cline{8-13}
  \end{tabular}
  }
\end{table*}

\subsubsection{Details of the E2E SA-ASR}

The AsrEncoder and AsrDecoder were configured to be almost 
the same with the encoder and decoder used
in the single-talker ASR system except that AsrDecoder
had one additional weight matrix to
feed $\bar{d}_n$ into AsrDecoder \cite{kanda2021end},
 and $\langle sc\rangle$ token
was added to the 4K-subword unit.
The SpeakerEncoder consisted of Res2Net as same with
 the d-vector extractor except 
that the final average-pooling layer was discarded.
 Finally, SpeakerDecoder consisted of 2 layers of Transformer \cite{vaswani2017attention},
 each of which had a multi-head attention with
 8 heads and a 2048-dim feed forward layer.
 We used a 80-dim log mel filterbank extracted every 10 msec as the input feature.

The E2E SA-ASR model was trained based on the two-stage pre-training 
with large-scale simulation followed by the final fine-tuning with the AMI-SDM training data.
%\begin{enumerate}
%\begin{itemize}
\begin{description}
\setlength{\leftskip}{-15pt}
\setlength{\itemindent}{-10pt}
\item {\bf The first stage pre-training}: 

   We optimized only
the ASR block as a speaker-agnostic multi-speaker ASR model
by setting $\bar{d}_n=0$. 
The multi-talker training data was simulated
based on the 75K data used for the single-talker ASR training.
Specifically, we randomly picked $N$ audio samples from 
the 75K-hour data
where $N$
was uniformly chosen from 1 to 5.
We assume these samples are from different speakers,
and simply mixed them by adding delay to each sample.
The delay amount was also randomly sampled under the constraints that
there was at least 0.5 sec of difference in the starting times of
the audio samples 
and that each audio sample had at least one overlapping region 
with another sample. 
After mixing the speech samples, speed perturbation \cite{ko2015audio} of 
0.9--1.1x was applied. 
We performed 425K training iterations with
32 GPUs, each of which consumed mini-batches of 24,000 frames.
We used Adam optimizer with
a linear decay learning rate schedule with a peak learning rate of 1e-3 
after 25K warm up iterations.

\item {\bf The second stage pre-training}: 

%\setlength{\leftskip}{-15pt}
%\item {In the second stage of the pre-training},

We optimized the E2E SA-ASR model by 
freezing the ASR block.
At this stage, we used 
the multi-talker simulation data based on the VoxCeleb corpus.
While the VoxCeleb corpus contains the speaker tag for each audio,
it does not contain the transcription.
Therefore, we first applied the single-talker ASR (before fine-tuning by AMI training data)
for each audio in the VoxCeleb
to generate the pseudo transcription.
Note that we used all data in the VoxCeleb corpus even though 
it contains many non-English contents. 
We then simulated roughly 9K hours of multi-talker recordings
by randomly mixing the audio from the VoxCeleb corpus
with the same setting used in the first-stage pre-training.
For each training sample,
we prepared $M={\rm uniform}(N, 8)$ speaker profiles where
$N$ profiles were extracted from the randomly selected utterances of 
the speakers in the audio sample, and $M-N$ profiles were extracted
from randomly selected utterances that was not included in the audio sample.
Network parameters in AsrEncoder and AsrDecoder were initialized 
by those of the first-stage pre-trained model.
Also, network parameters in SpeakerEncoder were
initialized by those of the speaker profile extractor.
We performed 110K training iterations with
32 GPUs, each of which consumed mini-batches of 16,000 frames.
We used Adam optimizer with
a linear decay learning rate schedule with a peak learning rate of 1e-3 
after 10K warm up iterations.

%Based on the pseudo transcription,

\item {\bf The final fine-training}: 
%\setlength{\leftskip}{-15pt}
%\item {In the second stage of the pre-training},

Finally, 
we optimized the entire E2E SA-ASR model by using 
the AMI-SDM training data. 
We segmented the training data based on the ``utterance-group'' determined in \cite{kanda2021large},
and used each segments for the training sample.
For each training sample,
we extracted $M={\rm uniform}(N, 8)$ speaker profiles where
$N$ speaker profiles were extracted for the relevant speakers and
$M-N$ speaker profiles were randomly extracted from the irrelevant speakers.
We performed 2,500 training iterations with
16 GPUs, each of which consumed mini-batches of 6,000 frames.
We used Adam optimizer with
a linear decay learning rate schedule starting from a learning rate of 1e-4.

%\end{itemize}
%\end{enumerate}
\end{description}

\section{Evaluation Results}

\subsection{Main results: modular v.s. joint systems}
We evaluated the four types of SA-ASR systems based on the AMI meeting corpus whose details are described in
Section \ref{sec:data}.
The cpWER and SCE for each system are shown in Table \ref{tab:main}.

% \underline{M1 v.s. M2:}
We first observed that
the M2 system significantly improved the cpWER over the M1 system for the
SDM recordings (from 28.6\% to 24.8\% in the development set and from 30.3\% to 28.4\% in the evaluation set).
We especially observed a significant improvement in deletion error, which
%should be attributed to the 
shows the
overlapping handling capability of the CSS module.
On the other hand, 
unexpectedly, we observed almost no improvement by the M2 system for IHM-MIX recordings.
This is possibly because we tuned all hyper-parameters based on the AMI-SDM development set,
and the CSS module was also trained for the noisy and reberverent speech.

We then observed that the J1 system significantly improved the cpWER from the modular systems
on most test conditions.
We also confirmed that the J2 system produced further improvement over the J1 system in all conditions.
For the SDM recordings, the J2 system achieved relatively 8.9--12.3\% better cpWER than the best modular system.
With the same setting, the J2 system achieved relatively 29.6--29.9\% better cpWER than the best modular system for the IHM-MIX recordings while we
think the difference could be reduced when we tuned the modular system towards IHM-MIX recordings.
In any case, we observed a significant cpWER improvement by the joint system over the modular ones.

Regarding the speaker counting, 
all systems showed less than 1 point of SCE,
which indicates the NME-based speaker counting worked very well.
%While we did not observe a significant difference among 
% four systems, t
The M2 system showed slightly better SCE in average, resulting from the leakage filtering module that effectively removed noise-only clusters.
Albeit, a better speaker counting accuracy was not the sufficient condition of a better cpWER, which also suggests the difficulty to tune the modular systems for overall performance.

\subsection{Errors attributed to speaker diarization}

To analyze the error distribution in more detail, 
we calculated the WER for J2 system. 
The results are shown in Table \ref{tab:wer}.
Note that the exact calculation of WER 
given multiple references and multiple hypotheses
is very computationally demanding procedure \cite{fiscus2006multiple}.
In our WER calculation, we restrict the search space by using the start and end time
of the VAD result. Due to this procedure, 
we are no longer able to compare the WER
between different systems, and we thus show the WER only for the J2 system.

We observed that the gaps between WER and cpWER are 2.9--4.0 points for SDM recordings
and  3.3--4.1 points for IHM-MIX recordings. 
These gaps can be attributed to the failure of speaker diarization\footnote{Due to the difference of the error calculation methods, the WER and cpWER numbers cannot be directly compared with each other. Therefore, we interpret the gap as an approximation of errors attributed to speaker diarization.}.
Note that, to the best of our knowledge, the state-of-the-art (SOTA) WER for the AMI-SDM recordings are 
18.4\% and 21.2\% for development set and evaluation set
with oracle VAD information, respectively \cite{kanda2021large}.
Also, the SOTA WER for IHM-MIX recordings are
13.5\% and 14.9\% for development set and evaluation set
with oracle VAD information, respectively \cite{kanda2021large}.
Our WER is very close to SOTA for SDM recordings and even better than SOTA for IHM-MIX recordings
even though our system is fully automated.

\begin{table}[t]
  \caption{The WER (\%) and cpWER (\%) for the J2 system. Note that we cannot compare the WER of different systems due to the method we used for WER calculation. Therefore, we show the result only for the J2 system to show the error attributed to speaker diarization. }
  \label{tab:wer}
%  \vspace{-3mm}
  \centering
 {
  \begin{tabular}{c|cc|cc}
   \hline 
Audio & \multicolumn{2}{c|}{dev} & \multicolumn{2}{c}{eval} \\
device            &   WER & cpWER & WER & cpWER \\
    \hline
IHM-MIX & 11.8  & 15.9 & 13.1 & 16.4 \\
SDM     & 18.6 & 22.6 &  22.0  & 24.9   \\ \hline
% IHM-MIX & 12.0  & 16.1 & 13.4 & 16.6 \\
% SDM     & 18.7 & 22.7&  22.3  &  26.0  \\ \hline
  \end{tabular}
  }
  \vspace{-3mm}
\end{table}

\subsection{Effect of fine-tuning}

To evaluate the importance of real training data,
we evaluated the SA-ASR systems before and after fine-tuning.
The results are shown in Table \ref{tab:fine-tuning}.
We found that the joint system was significantly worse than the modular system 
before fine-tuning while it became much better after the fine-tuning.
This result clearly shows the strength 
 of the joint system that can be optimized towards real data
 even if the data amount is
very small.
On the other hand, this result also suggests that 
the joint system could show deteriorated performance for the dataset in a different domain,
and the improvement of robustness to the domain mismatch can be an important future work.

\begin{table}[t]
  \caption{The cpWER (\%) for AMI SDM recording before and after fine-tuning of the ASR / E2E SA-ASR models. 
  }
  \label{tab:fine-tuning}
%  \vspace{-3mm}
  \centering
%  {\footnotesize
 % {\scriptsize
 {
  \begin{tabular}{cc|cc}
   \hline 
  System ID & Fine-tuning on  & \multicolumn{2}{c}{cpWER (\%)} \\
            & AMI       & dev & eval \\
    \hline
M2 & -  & 29.6 & 33.2 \\
M2 & $\surd$    & 24.8 & 28.4 \\ \hline
J2 & - & 54.1 & 52.3 \\
J2 & $\surd$   & {\bf 22.6} & {\bf 24.9}\\ \hline
% M2 & -  & 29.8 & 33.6 \\
% M2 & $\surd$    & 25.0 & 28.9\\ \hline
% J2 & - & 54.7 & 53.8 \\
% J2 & $\surd$   & {\bf 22.7} & {\bf 26.0}\\ \hline
  \end{tabular}
  }
  \vspace{-3mm}
\end{table}

\subsection{Gap between oracle and estimated information}

The J2 system is relying on two estimated information: 
the VAD results and the speaker cluster centroids from M1 system.
To diagnose the impact of the failure caused by these modules,
we evaluated the J2 system with oracle information.
Here, we used the utterance boundary information in the reference transcription
to get the oracle VAD result.
We used the same information to 
extract the speaker embedding from single-talker region for each speaker,
which was fed into the J2 system as the oracle speaker profile.

The results are shown in Table \ref{tab:oracle}.
We found that the failure in the VAD had more impact on the cpWER compared to
the failure in the speaker profile estimation.
For example, if we could use the oracle VAD information for IHM-recordings,
the cpWER became 14.8\% and 15.3\% for development and evaluation set, respectively.
These results are very close to the most oracle cpWER of 13.5\%\footnote{We found a mis-alignment of the IHM recordings and the transcriptions of the manual annotation (v1.6.2) of the session IB4002, which resulted in a degradation of cpWER. After manually fixing the incorrect alignment, the cpWER for the IHM development set became 12.6\%.} and 13.0\%
when we apply single-talker ASR for IHM recordings with the oracle utterance boundary (last row of Table \ref{tab:oracle}).
On the other hand, the difference of cpWERs between the oracle speaker profiles and
the estimated speaker profiles were relatively small, which indicates
 the quality of speaker clustering in the M1 system is sufficiently good.

\begin{table}[t]
  \caption{The comparison of oracle v.s. estimated information for J2 system.
  Note that the evaluation on the IHM recordings (last row) was evaluated by the single-talker
  ASR with oracle utterance boundary information.}
  \label{tab:oracle}
%  \vspace{-3mm}
  \centering
  %{\footnotesize
 % {\small
 % {\scriptsize
 {
  \begin{tabular}{c|cc|cc}
   \hline 
Audio  &  VAD & Speaker        & \multicolumn{2}{c}{cpWER (\%)} \\
device &      &  profile   & dev & eval \\
    \hline
 SDM     & automatic    & est. by M1    & 22.6  & 24.9 \\ % \hdashline
 SDM     & automatic    & {\bf oracle} & 22.4  & 24.7 \\
 SDM     & {\bf oracle} & est. by M1    & 20.8  & 23.2  \\
 SDM     & {\bf oracle} & {\bf oracle} & 20.7  & 22.8 \\ \hdashline
 IHM-MIX & automatic    & est. by M1    & 15.9 & 16.4\\ 
 IHM-MIX & automatic    & {\bf oracle} & 15.5 & 16.1 \\ 
 IHM-MIX & {\bf oracle} & est. by M1    & 14.8 & 15.3 \\  
 IHM-MIX & {\bf oracle} & {\bf oracle} & 14.5  & 15.0 \\  \hline
 IHM     & \multicolumn{2}{c|}{\bf oracle diarization} & 13.5$^3$ & 13.0\\  \hline
 % SDM     & automatic    & est. by M1    & 22.7  & 26.0 \\ % \hdashline
 % SDM     & automatic    & {\bf oracle} &  22.4 & 24.9 \\
 % SDM     & {\bf oracle} & est. by M1    & 20.8 & 24.3 \\
 % SDM     & {\bf oracle} & {\bf oracle} & 20.7 & 23.0 \\ \hdashline
 % IHM-MIX & automatic    & est. by M1    & 16.1 & 16.6 \\ 
 % IHM-MIX & automatic    & {\bf oracle} & 15.7 & 16.4 \\ 
 % IHM-MIX & {\bf oracle} & est. by M1    & 14.8 & 15.3 \\  
 % IHM-MIX & {\bf oracle} & {\bf oracle} & 14.5 & 15.0 \\  \hline
 % IHM     & \multicolumn{2}{c|}{\bf oracle diarization} & 13.5$^3$ & 13.0\\  \hline
  \end{tabular}
  \\
%  {\footnotesize $^\dagger$ The single-talker ASR was applied to IHM with oracle utterance boundary.}
% {\footnotesize $^\dagger$ The single-talker ASR was used.}
  }
  \vspace{-3mm}
\end{table}

\section{Conclusion}
In this paper, 
we presented our recent study on the comparison of modular 
and joint approaches for SA-ASR
%and joint approaches based on the E2E SA-ASR 
with real monaural recordings.  
We developed SOTA SA-ASR systems for
both modular and joint approaches by leveraging large-scale training data.
We also proposed a new pipeline that performs the E2E SA-ASR model after speaker clustering. 
Our evaluation on the monaural recordings in the AMI meeting corpus revealed 
that the joint system performs 8.9--29.9\% better in cpWER
compared to the modular system after fine-tuning by a small real data 
while the modular system performs better before such fine-tuning.  
We also conducted various error analyses to show the remaining issues for the monaural SA-ASR.

\bibliographystyle{IEEEbib}
\bibliography{mybib}

\end{document}